\documentclass[preprint,onecolumn,nofootinbib]{revtex4}
\pdfoutput=1
\usepackage[colorlinks=true,linkcolor=blue,urlcolor=blue,filecolor=black,citecolor=red,pdfstartview=FitV,pdftitle={},pdfsubject={},pdfkeywords={},pdfpagemode=None,bookmarksopen=true]{hyperref}
\usepackage{graphicx}%Include figure files
\usepackage{amsmath}
\usepackage{amsfonts}
\usepackage{amssymb,ulem}
\usepackage{color}%
\usepackage{CJK}
\usepackage{dcolumn}% Align table columns on decimal pointl
\newcommand{\f}{\begin{equation}}
\newcommand{\ff}{\end{equation}}
\newcommand{\fa}{\begin{eqnarray}}
\newcommand{\ffa}{\end{eqnarray}}

\begin{document}
\title{A simple holographic model for spontaneous breaking of translational symmetry}
\author{Wei-Jia Li$^{1}$}
\thanks{weijiali@dlut.edu.cn}
\author{Jian-Pin Wu$^{2}$}
\thanks{jianpinwu@yzu.edu.cn}
\affiliation{
$^1$ Institute of Theoretical Physics, School of Physics, Dalian
University of Technology, Dalian 116024, China\ \\
$^2$Center for Gravitation and Cosmology, College of Physical Science and Technology,
Yangzhou University, Yangzhou 225009, China
}
\begin{abstract}
It has been shown that holographic massive gravities can effectively realize spontaneous breaking of translational symmetry in homogenous manners. In this work, we consider a toy model of such category by adding a special gauge-axion coupling to the bulk action. Firstly, we identify the existence of spontaneous breaking of translations by the analysis on the UV expansion. In the absence of explicit breaking, the black hole solution is simply the same as  the Reissner-Nodstr\"{o}m(RN)  black holes, regardless of the non-trivial profile of the bulk axions. The  associated Goldstone modes exist only when the charge density is non-zero. Then, we investigate the optical conductivity both analytically as well as numercially. Our result perfectly agrees with that for a clean system, while the incoherent conductivity gets modified due to the symmetry breaking. The transverse Goldstone modes are dispersionless since the solution is dual to a \textit{liquid} state. Finally, the effect of momentum relaxation to the transverse modes is considered. In this case, the would-be massless modes are pinned at certain frequency, which is another key difference from the unbroken states.
\end{abstract}
\maketitle
\tableofcontents
\section{Introduction}
In most of real-word materials, the translational symmetry in spatial dimensions are inevitably broken both spontaneously and explicitly (In this paper, we call them SSB and ESB for short.)  due to the existence of
periodic lattice, striped orders, impurities, defects, etc. For crystalline states, the Goldstone bosons associated with the SSB of translations are usually called transverse and longitudinal phonons, all of which have linear dispersion relations and propagate freely at certain speed in the clean materials.  While, for liquid states with SSB of translations, there is only one longitudinal phonon since the shear stresses cannot be supported.

In strongly interacting electronic systems,  Goldstone modes and
electrons can be mightily coupled which gives rise to novel collective behaviors and exotic transport properties\cite{Hartnoll:1612,Bandurin:1509,Hartnoll:16mo}.  To have a deeper understanding on such patterns, building a framework beyond the conventional perturbative methods  has already become an important mission in condensed matter physics.

Holographic duality provides a tractable approach to the physics of strong correlated systems by mapping the many-body problems to classical gravity problems. Recently, some  holographic effective models for solid states that spontaneously break the translations have been constructed\cite{Amoretti:1611,Andrade:17081,Alberte:17082,Alberte:1711,Amoretti:1711,Amoretti:1712,Poovuttikul:1801,Baggioli:1805,Kuang:1808,Amoretti:1812}.  A common feature in these models is that the translations are spontaneously broken in a homogenous manner(For inhomogeneous realizations, one can refer to \cite{Rozali:1211,Donos:1303,Ling:1404,lili:1612,lili:1706,Jokela:1708,Donos:1801,Gouteraux:1803}). This significantly simplifies the calculations and makes it possible to dissect the key properties of the system, say, the transport, in analytic ways. Recently,
the homogeneous realization of phonons and pseudo-phonons has also been investigated in the field theory side\cite{musso}.

In this paper, we mainly consider a new simple holographic models which can realize the liquid states with SSB of the translations, by introducing a special gauge-axion coupling. In the absence of relaxation,  it is found that the background metric as well as the gauge field are exactly the same as the Reissner-Nodstr\"{o}m(RN)  black holes, while the profile of the bulk axions plays the role of the scalar condensate that breaks the translations. On top of this,  we investigate the imprints of the transverse Goldstone modes on the optical electric conductivity. The plan of this work is as follows: In section \ref{section2}, we construct the simplest holographic model with a  gauge-axion coupling and explain how the SSB of translations can happen in this model via analyzing the UV expansion of the bulk fields.  In section \ref{section3}, we compute the electric conductivity in the purely SSB pattern both analytically and numerically.   In section \ref{section4}, we consider the pinned modes in the presence of  relaxation. In section \ref{section6}, we conclude.

\section{Goldstone modes by gauge-axion coupling \label{section2}}
To break the translations in an isotropic and homogeneous manner, one needs to introduce massless axion fields with a bulk profile $\phi^I=k\,\delta^I_i\, x^i$.
Essentially, these scalar fields give the gravitons an effective mass, hence breaking the diffeomorphism invariance of the bulk theory\cite{vegh}. In this way, we can in principle construct a series of general effective holographic models with higher derivative terms\cite{WJL:1602}. Here, we will focus only on the following simple model with a special gauge-axion coupling:
\begin{equation}
\label{Model1}
\mathcal{S}\,=\,\,\,\int\,d^4x\,\sqrt{-g}\,\left(\,R\,-\,2\,\Lambda\,-\,\lambda X\,-\,\frac{1}{4}\,F^2\,-\,\frac{\mathcal{J}}{4}\,Tr[\mathcal{X}\,F^2]\,\right)
\end{equation}
where the U(1) gauge field $F_{\mu\nu}\equiv \nabla_\mu A_\nu-\nabla_\nu A_\mu$,
\begin{equation}
\label{Model1a}
Tr[\mathcal{X}\,F^2]\equiv {\mathcal{X}^\mu}_\nu {F^{\nu}}_{\rho} {F^\rho}_{\mu},\,\,\,\,\,\text{with} \,\,\, {\mathcal{X}^\mu}_\nu= \frac{1}{2}\sum_{I=x,y}\partial^\mu \phi^I \partial_\nu \phi^I,
\end{equation}
and $X\equiv Tr[\mathcal{X}]$. We require that $\lambda\ge 0$ for avoiding the ghost problem and a necessary condition $0\leq\mathcal{J}\leq2/3$ for unitarity and causality\cite{WJL:1602}. For convenience, we set the cosmological constant $\Lambda=-3$ which implies a normalized AdS radius. Then, if taking $\mathcal{J}=0$, it reduces to the simplest linear axion model.

From the action above, the covariant form of the equations of motion are given by
\begin{eqnarray}\label{a4}
\nabla_\mu\left[
F^{\mu\nu}-\frac{\mathcal{J}}{2}\left((\mathcal{X}F)^{\mu\nu}-(\mathcal{X}F)^{\nu\mu}\right)\right]=0\,,
\end{eqnarray}
\begin{eqnarray}\label{a5}
&&\nabla_\mu \Big[\lambda\nabla^\mu \phi^I+\frac{\mathcal{J}}{4}
{(F^2)^\mu}_\nu\nabla^\nu \phi^I\Big]=0\,,
\end{eqnarray}

and
\begin{eqnarray}\label{a6}
&&R_{\mu\nu}-\frac{1}{2}g_{\mu\nu}R-\frac{\lambda}{2}
\nabla_\mu \phi^I\nabla_\nu \phi^I-\frac{1}{2}\left(6-\frac\lambda 2\nabla_\sigma \phi^I \nabla^\sigma \phi^I\right)g_{\mu\nu}\\ \nonumber
&&=\frac{1}{2}
\Big({F_\mu}^{\sigma}F_{\nu\sigma}-\frac{1}{4}g_{\mu\nu}F_{\rho\sigma}F^{\rho\sigma}\Big)+\frac{\mathcal{J}}{4}\Big(\frac{1}{2}\nabla_{(\mu|} \phi^I\nabla_\sigma \phi^I{(F^2)^{\sigma}}_{|\nu)}\\ \nonumber
&&+F_{(\mu|\sigma}{(F\mathcal{X})^{\sigma}}_{|\nu)}+F_{(\mu|\sigma}{(\mathcal{X}F)^{\sigma}}_{|\nu)}-\frac{1}{2}g_{\mu\nu} Tr[\mathcal{X}F^2]\Big).
\end{eqnarray}
As is known that this model has the following isotropic charged black hole solutions:
\begin{eqnarray}\label{bansatz}
ds^2=-D(r)\,dt^2+B(r)\,dr^2+C(r)\, dx^idx_i, \ \ A_\mu=A_t (r)\,dt, \ \ \phi^I=(0,\,0,\,k\,x,\,k\,y),
\end{eqnarray}
where $i=2,3$ denotes the two spatial directions.  Choosing such radial coordinate $r$ that the AdS boundary is located at $r=0$, in the asymptotic region the background solution behaves like
\begin{align}\label{uvmetric}
&D(r)=\frac{1}{r^2}\left(1-d_{(3)}\,r^3+\dots\right),\nonumber\\
&B(r)=\frac{1}{r^2}(1+d_{(3)}\,r^3+\dots),\nonumber\\
&C(r)=\frac{1}{r^2},\nonumber\\
&A_t(r)=\mu-\rho \, r+\dots,
\end{align}
where the coefficient $d_{(3)}$ is associated with the energy density, $\mu$ and $\rho$ are the $U(1)$ chemical potential and the charge density in the boundary theory.

To explain why the SSB of translations can be realized in the holographic model (\ref{Model1a}), we firstly explain what role the profile of the scalars $\phi^i=k\,x^i$ plays in the following  two different cases. Without loss of generality, let us now assume that  $\phi^I$ depend on the full coordinates $x^\mu$. If we set $\lambda\neq0$ and $\mathcal{J}=0$ in (\ref{a5}), the asymptotic behavior of $\phi^I$ near the UV boundary is
\begin{eqnarray}\label{rebc}
\phi^I=\phi_{(0)}^I(t,x^i)+\phi_{(3)}^I(t,x^i)\,r^3+\dots.
\end{eqnarray}
Then, the $r-$independent term $\phi_{(0)}^i(t,x^i)$ dominates the second one, and hence plays the role of an external source that breaks the translations in the standard quantization. Obviously, the profile $\phi^i=k\,x^i=\phi_{(0)}^i$ means that the ESB of translations happens.  And this has already been widely investigated in previous holographic studies on momentum relaxation\cite{Andrade:1311,Gouteraux:1401,Kim:1501}.

Conversely, if we instead set $\lambda=0$ and $\mathcal{J}\neq0$, the scalars behave like
{\footnote{In practice, switching off the canonical kinetic term of the scalar fields theory means that the theory becomes strongly coupled at a relatively low energy scale. However, strong coupling is ubiquitous in field theories describing the low energy dynamics of systems with spontaneously broken translational symmetry.  That is to say the bulk EFT of our model should be valid only down to a certain radial scale. A parallel and detailed argument can be seen from \cite{Alberte:1711}. To achieve the expansion (\ref{ssbc}), we have also supposed that $A_t'(0)\sim const\neq0$, i.e., $\rho\neq0$.}
\begin{eqnarray}\label{ssbc}
\phi^I=\frac{\phi_{(-1)}^I(t,x^i)}{r}+\phi_{(0)}^I(t,x^i)+\dots.
\end{eqnarray}
In this case, the $r-$ independent term is subleading and corresponds to the expectation value of a dual operator $\mathcal{O}^I$.\footnote{See the holographic renormalization procedure in the Appendix.}
Now,  $\phi^i=k\,x^i$ should be interpreted as the expectation value $\langle\mathcal{O}^i\rangle\sim k\,x^i$ with a  vanishing source, i.e. $\phi_{(-1)}^i=0$. Since such a scalar condensate  $\langle\mathcal{O}^i\rangle$ is not uniform in $x^i$, the translational symmetry is broken spontaneously. The Nambu-Goldstone theorem claims that there should exist gapless excitations in the low energy description which are called Goldstone modes. With $\lambda=0$, the background solution is given by
\begin{align}
&D(r)=\frac{f(r)}{r^2}=\frac{1}{r^4B(r)},\\ \nonumber
&f(r)=1\,-\,\frac{r^3}{r_h^3}\,-\,\frac{\mu^2\,r^3}{4\,r_h}\left(1\,-\,\frac{r}{r_h}\right), \\ \nonumber
&C(r)=\frac{1}{r^2}\,, \qquad  A_t(r)=\mu-\rho\,r,\label{21}
\end{align}
where $r_h$ denotes the location of the horizon. Note that, unlike other massive gravity models, the background metric as well the gauge field do not depend on $k$. While the scalars can still have the non-trivial profile $\phi^i=k x^i$. Requiring the gauge field to be regular on the horizon gets $\mu=\rho\,r_h$. Finally, the Hawking temperature is given by
\begin{eqnarray}\label{22}
T=\frac{3}{4\pi r_h}-\frac{\mu^2 r_h}{16\pi},
\end{eqnarray}
which is also the same as the RN black holes. If we perturb the fields around the non-trivial solution like $\phi^I=k x^i+\chi^I$ and $A_\mu=A_t+a_{\mu}$, the axion-gauge term can be expanded as
\begin{eqnarray}\label{phlag}
\mathcal{L}_{\chi}&=&-\frac{\mathcal{J}r^4\rho^2}{8}\left[(\dot{\chi}^I)^2+({\chi^I}')^2\right]+\dots
\end{eqnarray}
where the dots represents the higher derivative terms associated with $\chi^I$ and $a_{\mu}$ interactions.  The leading term in the action is quadratic and the dynamics of the Goldstone modes in the dual boundary theory is encoded in the eom of $\chi^I$.  Note that this story happens only for finite density cases which is the similar case as the gapless sliding modes of charge density waves\cite{Gruner}, however in contrast to the acoustic phonons which do not carry $U(1)$ charges. For zero density case, the scalars $\chi^i$ will be decoupled from the other fluctuating fields in the bulk and will not affect the charge transport. In the next section, we will investigate the electric conductivity in the clean case.
\section{Electric conductivity}\label{section3}

We now turn to study small fluctuations around the background solution.
We denote $g_{\mu\nu}=\bar{g}_{\mu\nu}+\delta g_{\mu\nu}$, $A_\mu=\bar{A}_{\mu}+\delta A_{\mu}$ and $\phi^i=\bar{\phi}^i+\chi^i$,
where the quantities with bars are evaluated on the background, and introduce the time-dependent perturbations as follows
\begin{eqnarray}
&&
\delta A_\mu(t,r,x^i)=\int ^{+\infty}_{-\infty}\frac{d\omega d^{2}p_i}{(2\pi)^3}e^{-i \omega t+ip_ix^i}a_\mu(r),\nonumber\\
&&
\delta g_{\mu \nu}(t,r,x^i)=\int ^{+\infty}_{-\infty}\frac{d\omega d^{2}p_i}{(2\pi)^3}e^{-i \omega t+ip_ix^i}r^{-2}h_{\mu \nu}(r),\nonumber\\
&&
\chi^I(t,r,x^i)=\int ^{+\infty}_{-\infty}\frac{d\omega d^{2}p_i}{(2\pi)^3}e^{-i \omega t+ip_ix^i}\psi^I(r).
\end{eqnarray}\label{23}
To derive the conductivity, we focus on the homogeneous vector modes, namely setting all the momenta $p_i=0$. Since the system is isotropic, we only need to consider the $x$-component of the vector modes, namely $a_x$, $h_{tx}$, $h_{rx}$ and $\psi^{x}$.
The linearized Maxwell, scalar equation and Einstein equations read
\begin{eqnarray}
&&f a_x''-\frac{\mathcal{J}}{4} k^2 r^2 f a_x''-\frac{1}{4} \mathcal{J} k^2 r^2a_x' f'+a_x' f'-\frac{1}{2} \mathcal{J} k^2 r f a_x'-\frac{\mathcal{J} k^2 r^2 \omega ^2 a_x}{4f}+\frac{\omega ^2a_x}{f}\nonumber\\
&&+\frac{i}{4} \mathcal{J} k^2 \rho  r^2 \omega  h_{rx}-i \rho
   \omega  h_{rx}+\frac{1}{4} \mathcal{J} k^2 \rho  r^2 h_{tx}'-\rho h_{tx}'+\frac{1}{4} \mathcal{J} k^2 \rho  r h_{tx}'+\frac{1}{2} i \mathcal{J} k \rho  r
   \omega  \psi^x=0\,,\\ \label{24a}
&&f{\psi^x}''-\frac{2 i k \omega  a_x}{\rho  r}-k h_{rx} f'+f' {\psi^x}'-k fh_{rx}'-\frac{2 k f h_{rx}}{r}-\frac{i k \omega
   h_{tx}}{f}+\frac{2 f {\psi^x}'}{r}+\frac{\omega ^2 \psi^x}{f}=0,\\ \label{24b}
&&-\frac{i \mathcal{J} k^2 \rho  r^4 \omega  a_{x}}{4 f}+\frac{i \rho  r^2 \omega a_{x}}{f}+\frac{\omega ^2 h_{rx}}{f}-\frac{i \omega
   h_{tx}'}{f}-\frac{\mathcal{J}}{4}  k^2 \rho ^2 r^4 h_{rx}+\frac{1}{4} \mathcal{J} k\rho ^2 r^4 {\psi^x} '=0\,,\\ \label{24c}
&&f h_{tx}''+\frac{\mathcal{J}}{4}  k^2 \rho  r^4 f a_x'-\rho  r^2 f a_x'+i \omega
   f h_{rx}'-\frac{2 i \omega  f h_{rx}}{r}-\frac{2 f h_{tx}'}{r}-\frac{\mathcal{J}}{4}  k^2 \rho ^2 r^4
   h_{tx}-\frac{i \mathcal{J}}{4}  k \rho ^2 r^4 \omega  {\psi^x}=0\,,
   \nonumber\\
   \label{24d}
\end{eqnarray}
In this case, the mass of the spin-1 gravitons should be read as  ${\mathcal{M}(r)}^2=\mathcal{J}\frac{k^2\rho^2r^4}{4}$, which varies non-trivially along the radial direction\cite{WJL:1602}. In particular, its value in the UV and IR satisfies that
\begin{eqnarray}\label{nc}
 \mathcal{M}(0)=0,\,\,\,\,\text{and}\,\,\,\,\mathcal{M}(r_h)=\left(\mathcal{J}\frac{\pi k^2\mu^2}{s}\right)^{1/2}\sim \text{finite}.
\end{eqnarray}
As is pointed in the previous holographic study, this is exactly a condition for realizing the \textit{gapless} Goldstone bosons\cite{Alberte:17082}. The optical conductivity can be achieved numerically by setting the infalling boundary conditions at the horizon and solve the linearized equations of motion in the bulk. Since the electric current is a vector operator, the conductivity is sensitive to the transverse Goldstone modes but cannot mix with the longitudinal component which is a scalar mode. Then, one can directly read the information about the transverse modes from the conductivity.\footnote{Peaks in the spectral function is associated with the low-lying quasi-normal modes on the complex plane. Then, the dispersion relation of the Goldstone modes can be identified by the motion of the peaks. We would like to thank M. Baggioli for pointing out this.}

Without explicit breaking, the optical conductivity of a \textit{relativistic} system in the hydrodynamic limit can be written as\cite{Hartnoll:1612}
\begin{eqnarray}\label{dcfull}
\sigma(\omega)\xrightarrow{\omega\rightarrow0}\sigma_0+\frac{\chi_{JP}^2}{\chi_{PP}}\frac{i}{\omega}
\end{eqnarray}
where $\chi_{JP}=\rho$, $\chi_{PP}=\epsilon+P$ in our case and the finite part $\sigma_0$ is the incoherent conductivity that is theory-dependent and irrelevant to the momentum relaxation. Unlike the DC conductivity,  the incoherent one can always be achieved via the membrane paradigm, hence is  UV insensitive from the RG perspective. In our model, it can be obtained similarly as in \cite{Amoretti:1711,Amoretti:1712}:
\begin{eqnarray}\label{dc}
\sigma_0=\left(\frac{sT}{sT+\mu\rho}\right)^2\left(1-\mathcal{J}\frac{\pi k^2}{s}\right),
\end{eqnarray}
where the thermodynamic relation $sT+\mu\rho=\epsilon+P$ has been applied. Eq.(\ref{dcfull}) means that the real part of conductivity has a delta infinity at zero frequency in the purely SSB pattern due to the absence of momentum relaxation.

In Fig.\ref{sigma}, we show that the numeric result of our holographic model with $\lambda=0$ agrees with (\ref{dcfull}) and (\ref{dc}) very well. This again indicates that the background profile of the scalars ${\bar{\phi}}^i=kx^i$ should be interpreted as breaking the translational symmetry  spontaneously rather than explicitly. And the Goldstone modes  contribute the correction term in (\ref{dc}) to the incoherent conductivity that is controlled by the parameters $\mathcal{J}$ and $k$.

The Goldstone modes associated with the SSB of translations in crystals are usually called phonons. Nevertheless, the  transverse gapless modes in this model are not phonon like. For transverse and longitudinal phonons, they have the linear dispersion relations $\omega\sim v_{T,L}\,p$ with finite sound speeds $v_{T,L}$. Setting $p_y=p\neq 0$ and repeat the numerical calculations(To do so, we adopt the gauge invariant formulation of the bulk fields like in \cite{baggioli:1411}, and solve three coupled equations of the shear modes numerically.), we find that the infinite peak does not move away from $\omega=0$ as the momentum $p$ is increased. In Fig.\ref{sigma2}, we show the imaginary part of the conductivity with the varying momentum.

Therefore, the Goldstone modes are not dispersive, which means there is no freely propagating transverse phonons.  This can also be understood in another way: For phonons, their sound speeds are related to the shear modulus $\mathcal{G}$ and the  bulk modulus $\mathcal{K}$ by\cite{Kadanoff:1963}
\begin{eqnarray}\label{sspeeds}
v_T^2=\frac{\mathcal{G}}{\chi_{PP}},\,\,\,\,\, v_L^2=\frac{\mathcal{G}+\mathcal{K}}{\chi_{PP}}.
\end{eqnarray}
One can however check that in our model the $\mathcal{J}$ coupling does not contribute a finite mass term to the spin-2 gravitons $\delta g_{xy}$. Then, we conclude that the shear viscosity obeys the KSS bound $\eta=\frac{s}{4\pi}$ and the shear modulus $\mathcal{G}=0$ which is the case for a strongly coupled fluid instead of a solid. First relation in (\ref{sspeeds}) implies that $v_T=0$, the transverse modes are not phonon like.  This does not conflict with the common sense that there is no transverse phonons in a fluid due to the lack of  shear stress.  Therefore, this holographic model provides a low energy description for the gapless Goldstone modes coupled with a \textit{conformal fluid}. In the next section, we will further study the optical conductivity in the presence of the explicit source that breaks the translations. The numeric result of the holographic model captures another key feature of the SSB of translations which is called pinning effect.

%%%%%%%%%%%%%%%%%%%%%%%%%%%%%%
%%%%%%%%%%%%%%%%%%%%%%%%%%%%%%
\begin{figure}
\center{
\includegraphics[scale=0.6]{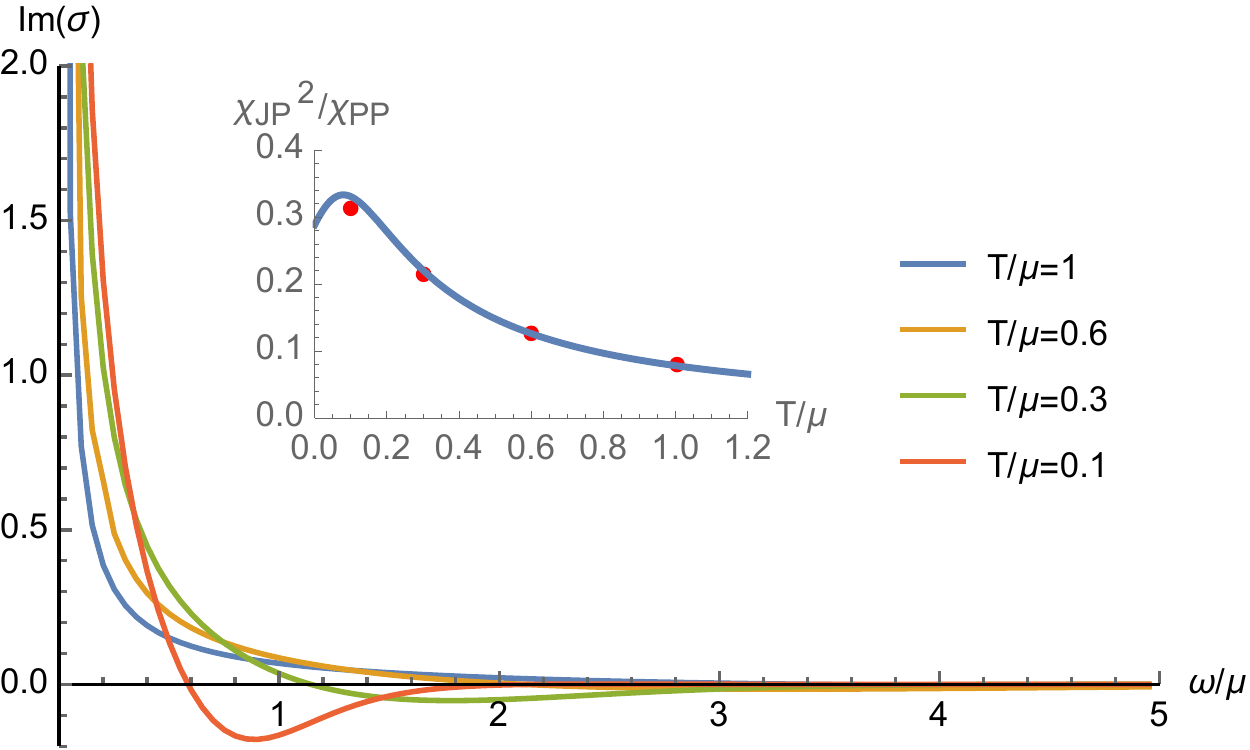}\ \hspace{0.8cm}
\includegraphics[scale=0.55]{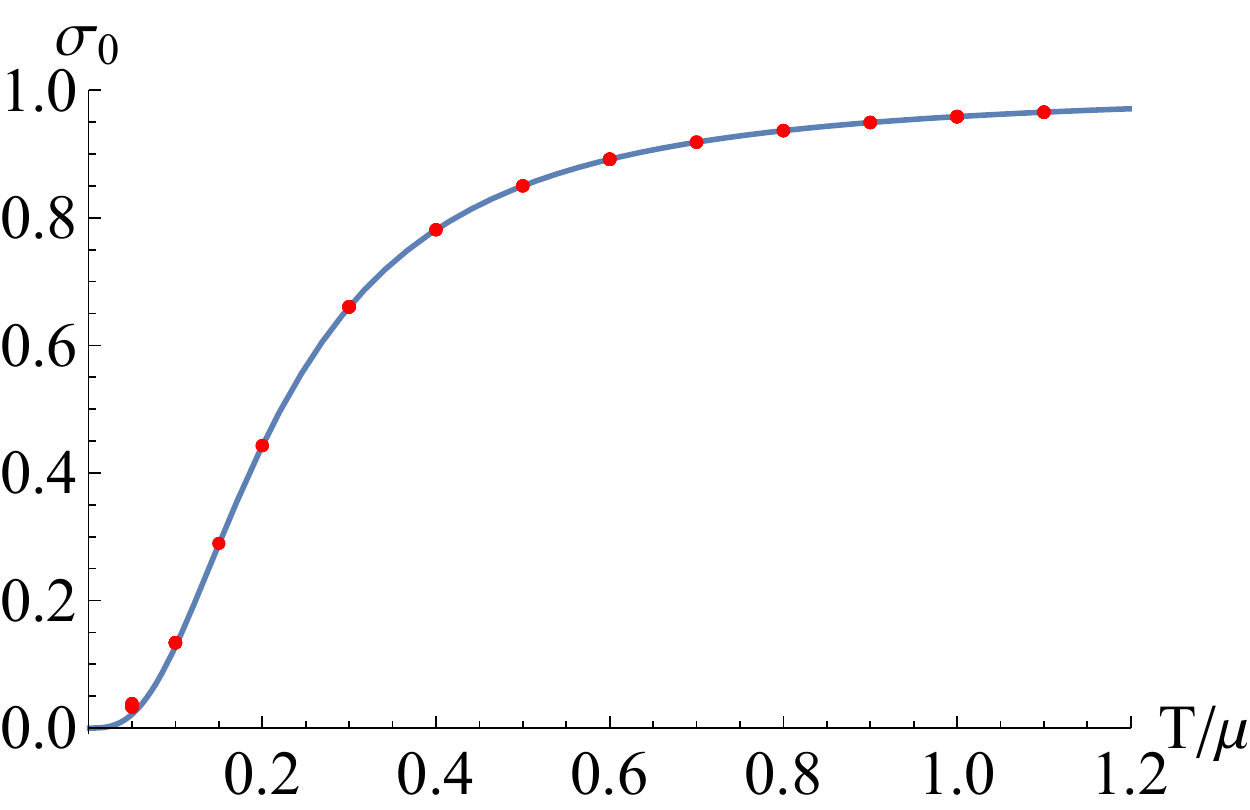}\ \hspace{0.8cm}\ \\
\caption{\label{sigma} Left plot: $\text{Im}(\sigma)$ as the function of $\omega/\mu$ for different temperature.
The inset plot shows the Drude weight, $\chi_{JP}^2/\chi_{PP}$, as the function of $T/\mu$. Blue line is the analytical result obtained from Eq.\eqref{dcfull}
and the red dots are the numerical result.
Right plot: The incoherent  conductivity $\sigma_0$ as the function of $T/\mu$.
Blue line is the analytical result obtained from Eq.\eqref{dc}
and the red dots are the numerical result.
Here, we have set $\mathcal{J}=1/3$ and $k/\mu=1$.}}
\end{figure}
%%%%%%%%%%%%%%%%%%%%%%
%%%%%%%%%%%%%%%%%%%%%%
\begin{figure}
\center{
\includegraphics[scale=0.60]{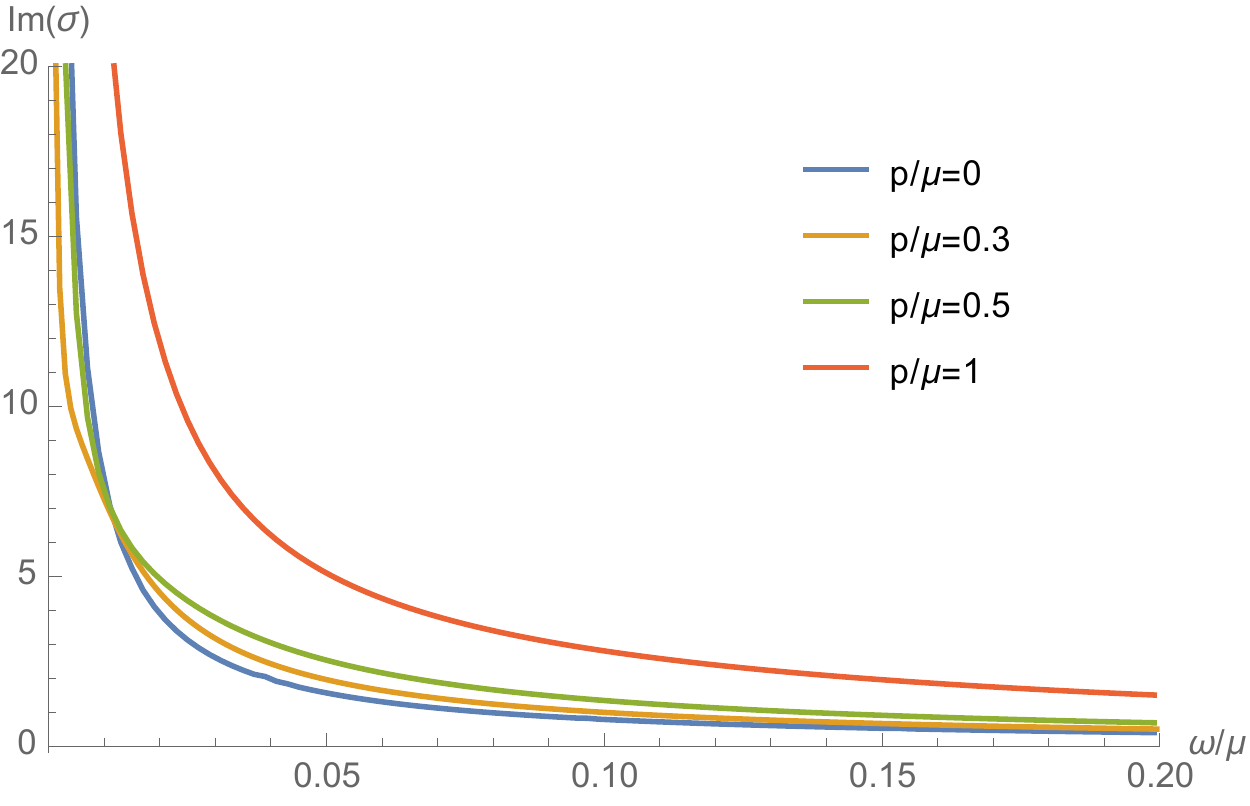}\ \hspace{0.65cm}\ \\
\caption{\label{sigma2} $\text{Im}(\sigma)$ as the function of $\omega/\mu$ for different momenta.
Here, we have set $\mathcal{J}=1/3$, $k/\mu=1$ and $T/\mu=1$. For all the cases, there is always an infinite pole standing still at zero frequency.}}
\end{figure}
%%%%%%%%%%%%%%%%%%%%%%
%%%%%%%%%%%%%%%%%%%%%%

\section{Pinning effect}\label{section4}
Now, we consider how the peak of the goldstone bosons moves in the presence of ESB. According to the UV analysis in section \ref{section2}, such a  pinning effect can be realized  by setting a non-zero value of $\lambda$, or the external source equivalently. In consequence, the infinite delta at zero frequency should be removed and there will be a sharp but finite peak at a certain frequency (called pinning frequency) in the optical conductivity.

When $\lambda\neq0$, the blackfactor $f(r)$ becomes
\begin{align}
&f(r)=1\,-\,\frac{r^3}{r_h^3}\,-\,\left(\frac{\lambda k^2 r^2}{2}+\frac{\mu^2\,r^3}{4\,r_h}\right)\,\left(1\,-\,\frac{r}{r_h}\right).
\end{align}
And the linearized equations of motion:

\begin{eqnarray}
&&f a_x''-\frac{\mathcal{J}}{4} k^2 r^2 f a_x''-\frac{1}{4} \mathcal{J} k^2 r^2a_x' f'+a_x' f'-\frac{1}{2} \mathcal{J} k^2 r f a_x'-\frac{\mathcal{J} k^2 r^2 \omega ^2 a_x}{4f}+\frac{\omega ^2a_x}{f}\nonumber\\
&&+\frac{i}{4} \mathcal{J} k^2 \rho  r^2 \omega  h_{rx}-i \rho
   \omega  h_{rx}+\frac{1}{4} \mathcal{J} k^2 \rho  r^2 h_{tx}'-\rho h_{tx}'+\frac{1}{4} \mathcal{J} k^2 \rho  r h_{tx}'+\frac{1}{2} i \mathcal{J} k \rho  r
   \omega  \psi^x=0\,,\\
&&f{\psi^x}''-\frac{2 i k \omega  a_x}{\rho  r}-k h_{rx} f'+f' {\psi^x}'-k fh_{rx}'-\frac{2 k f h_{rx}}{r}-\frac{i k \omega
   h_{tx}}{f}+\frac{2 f {\psi^x}'}{r}+\frac{\omega ^2 \psi^x}{f}
   -\frac{4\lambda k  f'  h_{rx}}{\mathcal{J} \rho ^2 r^4}
   \nonumber\\
 &&
 +\frac{4 \lambda  f'{\psi^x}'}{\mathcal{J} \rho ^2 r^4}-\frac{4\lambda k  f h_{rx}'}{J \rho ^2 r^4}+\frac{8 \lambda k
     f h_{rx}}{J \rho ^2 r^5}-\frac{4 \lambda i k \omega h_{tx}}{\mathcal{J} \rho ^2 r^4 f}-\frac{8 \lambda  f {\psi^x} '}{J \rho ^2r^5}+\frac{4 \lambda  f{\psi^x} ''}{J \rho ^2 r^4}+\frac{4 \lambda  \omega ^2 {\psi^x}}{J \rho ^2 r^4 f}=0\,,\\
&&\lambda k{\psi^x}'-\lambda k^2h_{rx}-\frac{i \mathcal{J} k^2 \rho  r^4 \omega  a_{x}}{4 f}+\frac{i \rho  r^2 \omega a_{x}}{f}+\frac{\omega ^2 h_{rx}}{f}-\frac{i \omega
   h_{tx}'}{f}-\frac{\mathcal{J}}{4}  k^2 \rho ^2 r^4 h_{rx}+\frac{1}{4} \mathcal{J} k\rho ^2 r^4 {\psi^x} '=0\,, \nonumber\\
 &&
 \
 \\
&&f h_{tx}''+\frac{\mathcal{J}}{4}  k^2 \rho  r^4 f a_x'-\rho  r^2 f a_x'+i \omega
   f h_{rx}'-\frac{2 i \omega  f h_{rx}}{r}-\frac{2 f h_{tx}'}{r}-\frac{\mathcal{J}}{4}  k^2 \rho ^2 r^4
   h_{tx}-\frac{i \mathcal{J}}{4}  k \rho ^2 r^4 \omega  {\psi^x}\nonumber\\
 &&-\lambda k^2h_{tx}-i\lambda k\omega \psi^{x}=0\,,
\end{eqnarray}
From the equations, the effective mass of the spin-1 gravitons should be identified as $\mathcal{M}(r)^2\equiv k^2\left(\lambda+\frac{\mathcal{J}  \mu^2 r^2}{4}\right)$.  With $\lambda\neq 0$, the DC conductivity can be directly computed by the membrane paradigm \cite{WJL:1602}
\begin{equation}\label{dcre}
\sigma_{DC}=\left(1-\mathcal{J}\frac{\pi k^2}{s}\right)+\left(1-\mathcal{J}\frac{\pi k^2}{s}\right)^2\frac{\mu^2}{\mathcal{M}(r_h)^2},\,\,\,\,
\end{equation}
The gapped  modes can be identified with the peaks in the optical conductivity in the two plots of Fig.\ref{fig-AC-diff-T}. As is expected, the pinning frequency becomes higher as the increase of $\lambda$, i.e, the rate of momentum relaxation.  However, the relation between $\omega_0/\mu$ and $\lambda$ exhibits a peculiar scaling that is different from the  Gell-Mann-Okubo relation. In addition, it is obvious to see from the left plot of Fig.\ref{fig-AC-diff-T2} that the pinning effect introduces a mechanism of transition from a metallic state(when $\mathcal{J}=0$)  to an insulating state, which is quite similar as what happens in a doped Mott insulator\cite{tomas:1710}. Even though, the commensurability  effect is in general absent in the holographic systems with homogeneity\cite{tomas:1512}.

The quantitive relation between $\lambda$ and the relaxation rate can be in principle checked by a full analysis on the quasi-normal modes of the black hole like in \cite{Alberte:17082,Baggioli:1807,Baggioli:1808}, which is however not a target in this work.  Now, we consider how the propagation of the transverse modes would change when we vary the value of $\lambda$. We turn on $p\neq 0$ and obtain the optical conductivity with finite momentum in the right plot of Fig.\ref{fig-AC-diff-T2}.
The numerical result shows that the peak of gapped modes becomes milder as increasing the momentum. However, these modes are still dispersionless, in contrast to the solid holographic massive gravity model\cite{Alberte:17082,baggioli:1411}.

\begin{figure}
\center{
\includegraphics[scale=0.6]{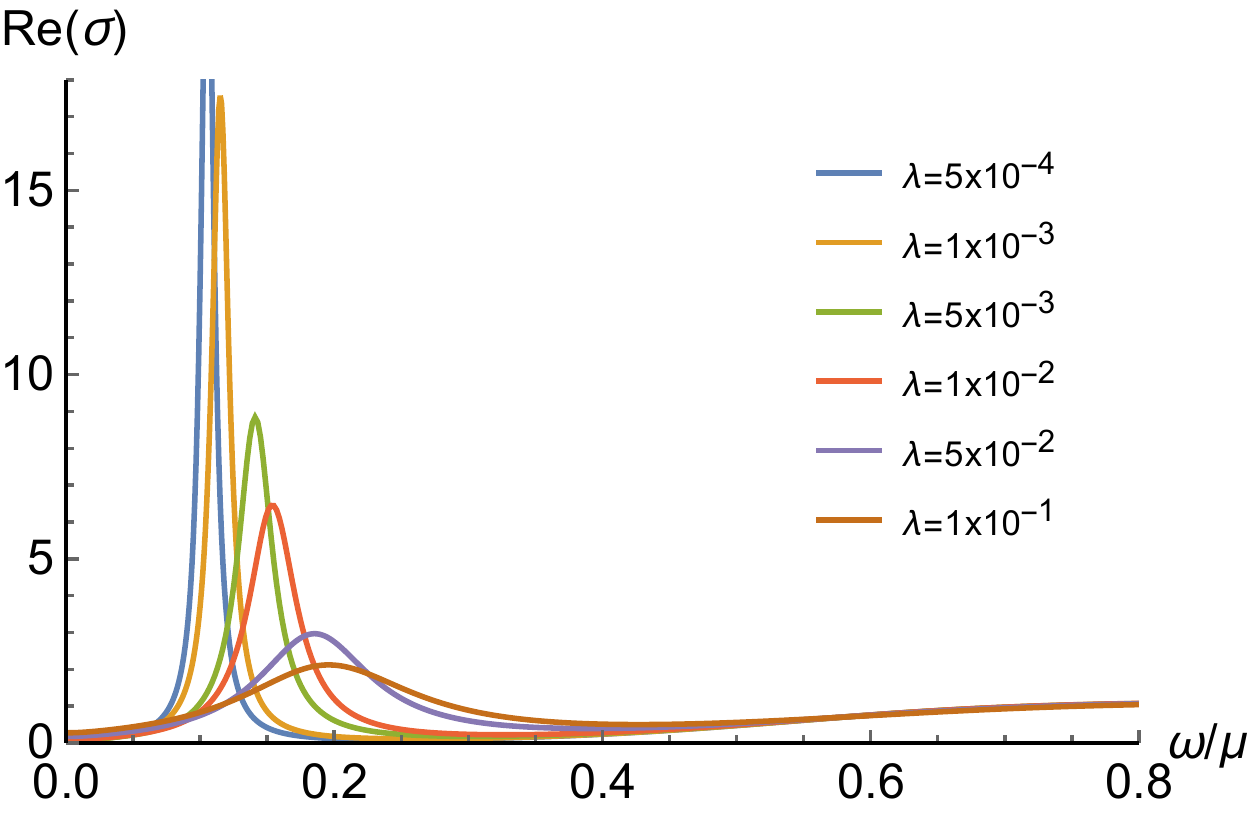}\ \hspace{0.8cm}
\includegraphics[scale=0.6]{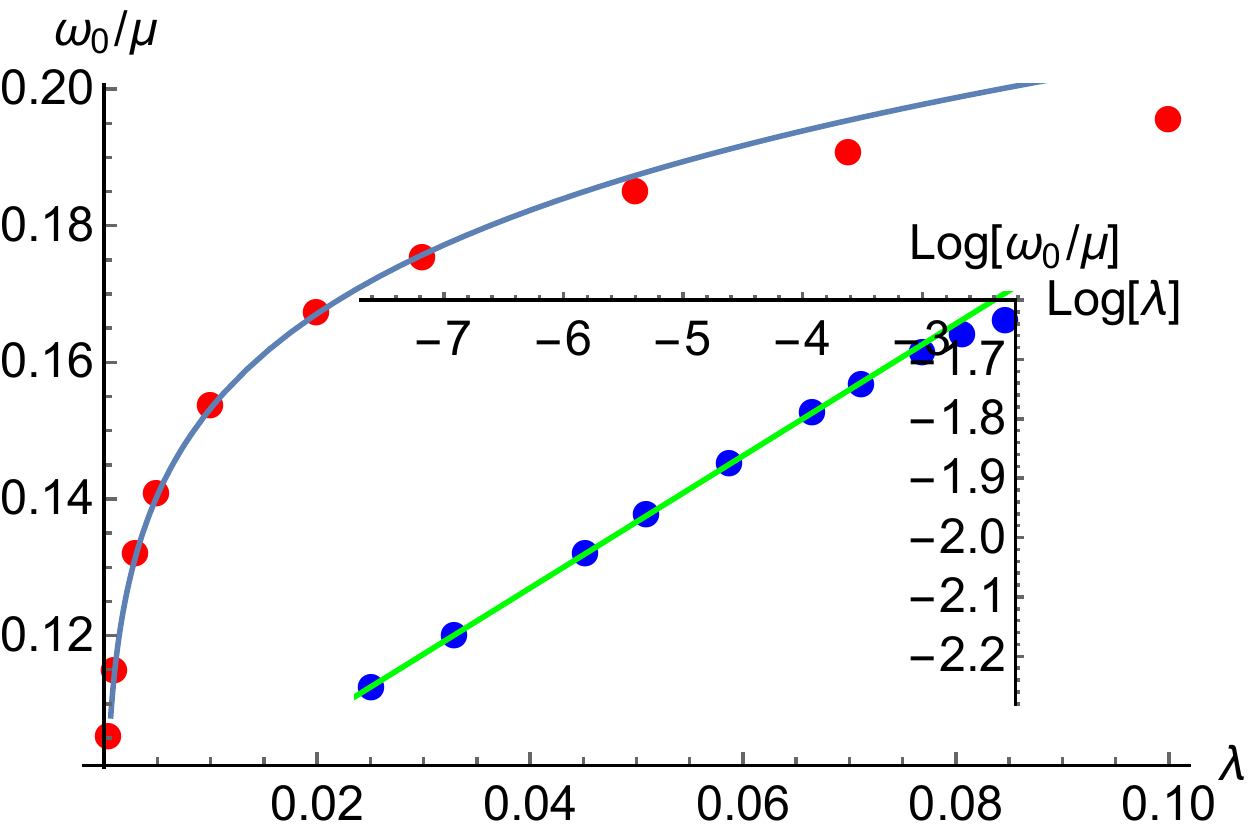}\ \\
\caption{\label{fig-AC-diff-T} Left plot: $\text{Re}(\sigma)$ as the function of $\omega/\mu$ for different values of $\lambda$. Right plot: The pinning frequency as the function of $\lambda$. The solid lines are the fitting results which show that $\omega_0/\mu\approx 0.272532 \,\lambda^{0.125} $ in the region $\lambda\in [5\times10^{-4},3\times10^{-2}]$.  Here, we have fixed $\mathcal{J}=1/3$, $k/\mu=1$ and $T/\mu=0.005$.
 }}
\end{figure}

\begin{figure}
\center{
\includegraphics[scale=0.6]{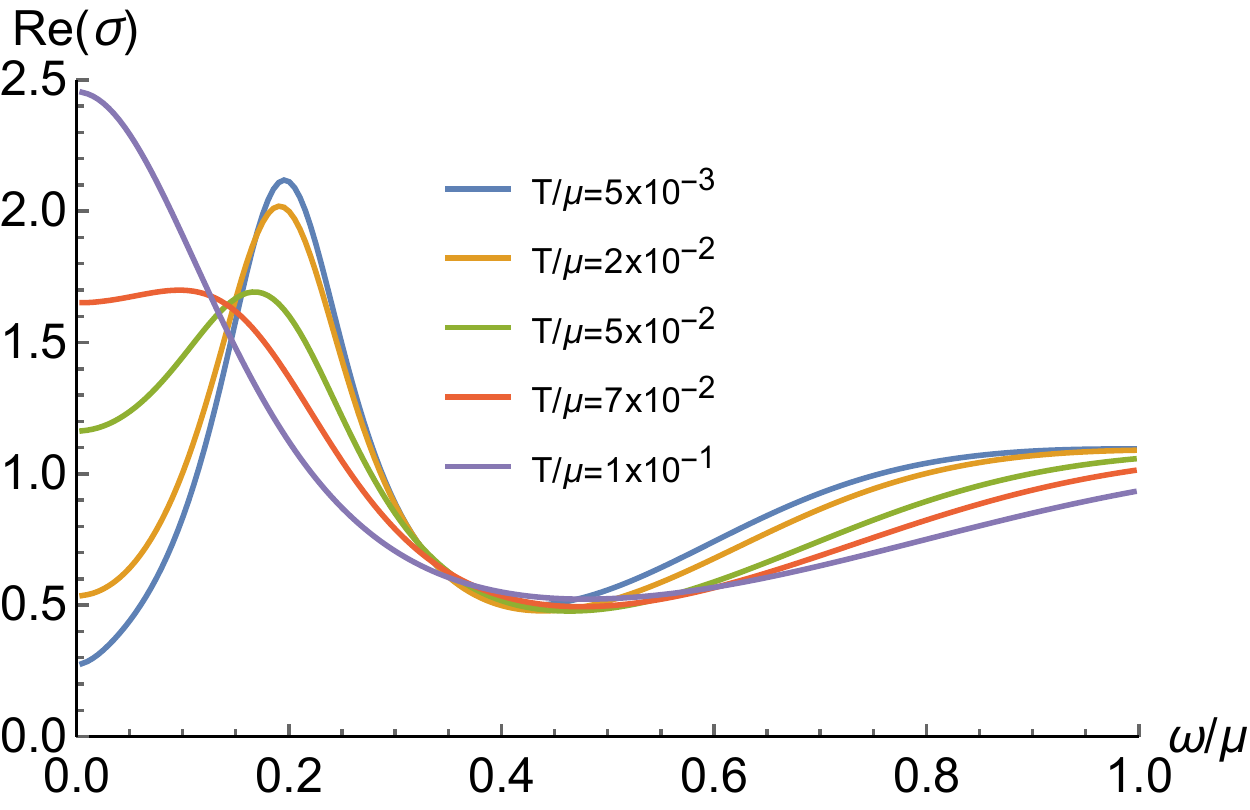}\ \hspace{0.8cm}
\includegraphics[scale=0.6]{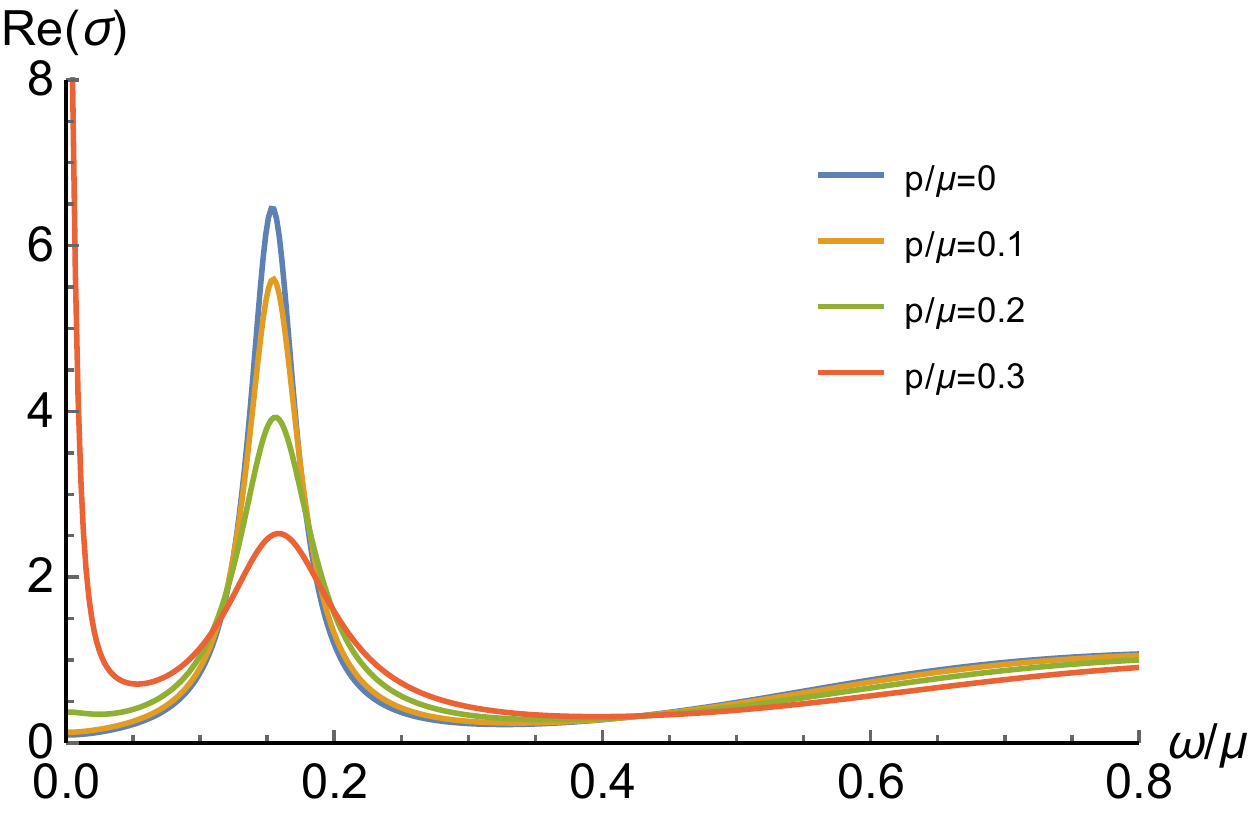}\\\
\caption{\label{fig-AC-diff-T2} Left plot: $\text{Re}(\sigma)$ as the function of $\omega_0/\mu$ for $\lambda=0.1$ and various values of $T/\mu$. It shows that $d\sigma_{dc}/dT>0$. Right plot: The optical conductivity for $\lambda=0.01$, $T/\mu=0.005$ and various values of $p/\mu$. Note that the pinned modes are dispersionless which is the same as the purely SSB pattern. However, there appears a gapless peak again for large values of $p/\mu$. The underlying physics still remains
to be revealed in future. In both plots, we have fixed $\mathcal{J}=1/3$, $k/\mu=1$.}}
\end{figure}

\section{Conclusion and outlook} \label{section6}
In this paper, we introduce a simple holographic model that can realize both the spontaneous and explicit breaking of the translational symmetry in the dual field theory.  In this model, the SSB is induced by a gauge-axion coupling $\mathcal{J}Tr[\mathcal{X}F^2]$, while the ESB can be realized by turning on the linear axion term $\lambda X$.

When we turn off the external scalar source by setting $\lambda=0$, the condensate of the dual operators that breaks the translations should be identified as the bulk profile of the axions, via the UV analysis.  In this case, the background metric and gauge field is the same as the RN black holes. And the dynamics of the transverse Goldstone modes is encoded in the eoms of the spatial components of axions. Our numeric result of electric conductivity matches with that of a fluid with SSB of translations. We then turn on the explicit source to see its pinning effect on the Goldstone modes. It is found that the pinning frequency becomes higher as we increase the value of $\lambda$. And the gapped modes are still dispersionless.

In this short paper, the analysis on the bulk mode is lacking.  In fluids, there exists longitudinal phonons whose speed is related to both of the shear and bulk modulus.  Then, the second relation in (\ref{sspeeds}) can be checked by studying the coupled spin-0 fluctuations, which is more complicated.\footnote{Since the $\mathcal{J}$ coupling does not affect the background, we cannot directly express the bulk modulus $\mathcal{K}$ in terms of the background quantities as in \cite{Baggioli:1805,Baggioli:up}.} We will leave this for future work\cite{WJL:up}.

Our model can be generalized, including the higher derivative terms like $\sum_{n=2}^{\infty}Tr[\mathcal{X}^nF^2]$. One can, however, check that such terms do not change the UV expansion (\ref{ssbc}), hence will not change the story a lot, albeit further modifications on the incoherent conductivity. One can also consider another class of gauge-axion couplings, for instance, $K\,Tr[X]F^2 $\cite{WJL:1602}.  This term will change the background solution and gives the system a non-zero shear modulus when $\rho\neq0$. Then the dual system may be interpreted as some kind of ``electronic crystals", whose impacts on the  transport or elastic properties are also worth studying.
In \cite{Lindgren:2015lia,Caldarelli:2016nni}, a general framework has been developed for computing the holographic 2-point function
and the corresponding conductivities dual to a broad class of Einstein-Maxwell-Axion-Dilaton theories.
We can generalize this study to include the higher derivative axion terms and see what novel phenomena emerge.
In future, we will study this issue following the line in \cite{Lindgren:2015lia,Caldarelli:2016nni}.

\acknowledgments

We are particularly grateful to M. Baggioli for many stimulating discussions, sharing us the Mathematica code for the optical conductivity and reading a preliminary version of the draft.
We also thank C. Niu for helpful discussions on the numerics.
This work is supported by the Natural Science Foundation of China under Grant No.11775036, 11847313 and Fundamental Research Funds for the Central Universities No.DUT 16 RC(3)097.
WJL also would like to thank Shanghai University and Shao-Feng Wu for the warm hospitality during the completion of this work.

\begin{appendix}
\section*{Appendix: Holographic renormalization}
To clarify the identification of the coefficients in the UV expansion (9), we compute the one-point correlator of the dual scalar operator by performing the holographic renormalization procudure\cite{Skenderis,Amoretti:1711} for the purely SSB case. For this purpose, we only need to focus on the gauge-axion coupling that describes the dynamics of broken phase,
\begin{align}\label{gacoupling}
S^{\text{SSB}}&=-\frac{\mathcal{J}}{8}\int d^4 x\sqrt{-g} \left(\partial^\mu \phi^I \partial_\nu \phi^I F^{\nu\rho}F_{\rho \mu} \right)\nonumber\\
&=-\frac{\mathcal{J}}{4}\int_{r=\epsilon} d^3 x\sqrt{-g} g^{rr}\left(\phi^I \partial_\nu \phi^I F^{\nu\rho}F_{\rho r}-\partial_r \phi^I \partial_\nu \phi^I F^{\nu\rho}A_{\rho}\right)+\text{bulk terms},
\end{align}
where $\epsilon$ is the UV cut-off, the bulk terms can be eliminated by using the the on-shall condition and the eoms of the bulk fields. In order to calculate the one-point correlation, $\langle\mathcal{O}^I\rangle$, we derive the renormalized action up to linear order in the perturbations.  Consider the fluctuations around the background solution(\ref{bansatz}):
\begin{align}\label{perturbations}
&g_{\mu\nu}=\bar{g}_{\mu\nu}+\delta g_{\mu\nu},\nonumber\\
&A_{\mu}=\bar{A}_{\mu}+\delta A_{\mu},\nonumber\\
&\phi^I=\bar{\phi}^I+\delta \phi^{I}.
\end{align}
Here, $\bar{g}_{\mu\nu},\bar{A}_{\mu},\bar{\phi}^I$ denote the background fields and $\delta g_{\mu\nu},\delta A_{\mu},\delta \phi^{I}$ denote the fluctuations.
Similar as the UV expansion (\ref{uvmetric}) and (\ref{ssbc}), the fluctuations in the asymptotic region behave as:
\begin{align}\label{perturbationsUV}
&\delta g_{\mu\nu}=\frac{1}{r^2}\left(\delta g_{\mu\nu}^{(0)}\,+\,\delta g_{\mu\nu}^{(1)}\,r\,+\,\dots\right),\nonumber\\
&\delta A_{\mu}=\delta A_{\mu}^{(0)}\,+\,\delta A_{\mu}^{(1)}\,r+\dots,\nonumber\\
&\delta \phi^{I}=\frac{\delta \phi_{(s)}^I}{r}\,+\,\delta \phi_{(v)}^I+\dots.
\end{align}
Inserting this into (\ref{gacoupling}) gives that
\begin{align}\label{regularv1}
\delta S^{\text{SSB}}_{\text{reg}\,(1)}&=-\frac{\mathcal{J}}{4}\int_{r=\epsilon} d^3 x\sqrt{-\bar{g}} \bar{g}^{rr} \bar{\phi}^I \partial_r \delta\phi^I \bar{F}^{rt}\bar{F}_{t r},\nonumber\\
&=\frac{\mathcal{J}\rho^2}{4}\int_{r=\epsilon} d^3 x\left(\frac{\bar{\phi}_{(s)}^I \delta\phi_{(s)}^I}{r}-\bar{\phi}_{(v)}^I \delta\phi_{(s)}^I+\dots\right).
\end{align}
In the first line above, only one term associated with $\delta \phi^I$ survives when we expand the on-shell action to linear level because of the fact that the background value of $\phi^I$ does not depend on $t$ or $r$. Then, all the terms containing $\delta g_{\mu\nu}$ or $\delta A_{\mu}$ should be vanishing. Moreover, only the first term in the second line is divergent as $\epsilon\rightarrow0$, which can however be removed by subtracting an additional boundary counter-term:
\begin{align}\label{regular}
S_{\text{c.t.}}^{\text{SSB}}=\frac{\mathcal{J}}{16}\int_{r=\epsilon} d^3 x\sqrt{-\gamma} (\phi^I)^2F^2.
\end{align}
 The last term in (\ref{regularv1}) is finite, which corresponds to the coupling term of the scalar operator $\mathcal{O}^I$ in the field theory side. Then, one can easily read that
\begin{align}\label{vev}
\langle\mathcal{O}^I\rangle=\frac{\mathcal{J}\rho^2}{4}\bar{\phi}_{(v)}^I.
\end{align}
Note that $\bar{\phi}_{(v)}^I$ was denoted as $\phi^I_{(0)}$ in the UV expansion (\ref{ssbc}) in the main text. We thus conclude that the coefficient $\phi_{(0)}^I$
indeed corresponds to the expectation value of the operator $\mathcal{O}^I$, as expected. Hence, the background solution $\phi^I=k\delta_i^I x^i$ should be interpreted as a translational order, $\langle\mathcal{O}^i\rangle=\mathcal{J}\rho^2\phi^I_{(0)}/4\sim k\,x^i$.
\end{appendix}


\begin{thebibliography}{99}
 \bibitem{Hartnoll:1612}
S. A. Hartnoll, A. Lucas, and S. Sachdev, ``Holographic quantum matter", MIT press.
 \bibitem{Bandurin:1509}
D. Bandurin \textit{et.al}, ``Negative local resistance caused by viscous electron backflow in graphene", Science 351 (2016), no. 6277 1055-1058, [arXiv:1509.04165].
 \bibitem{Hartnoll:16mo}
 L. Delacr\'etaz, B. Gout\'eraux, S. Hartnoll, A. Karlsson, SciPost Phys. 3, 025 (2017), arXiv:1612.04381 [cond-mat.str-el]; L. Delacr\'etaz, B. Gout\'eraux, S. Hartnoll, A. Karlsson, Phys. Rev. B 96, 195128 (2017),  arXiv:1702.05104 [cond-mat.str-el]
  \bibitem{Esposito:1708}
A. Esposito, S. Garcia-Saenz, A. Nicolis, R. Penco, ``Conformal solids and holography", 	JHEP 1712 (2017) 113,  arXiv:1708.09391[hep-th].
 \bibitem{Amoretti:1611}
A. Amoretti, D. Are\'an, R. Argurio, D. Musso, L. A. P. Zayas, ``A holographic perspective on phonons and pseudo-phonons", JHEP 05 (2017) 051,  arXiv:1611.09344 [hep-th].

 \bibitem{Andrade:17081}
T. Andrade, M. Baggioli, A. Krikun, N. Poovuttikul, ``Pinning of longitudinal phonons in holographic spontaneous helices", JHEP 1802 (2018) 085,  arXiv:1708.08306 [hep-th].
 \bibitem{Alberte:17082}
L. Alberte, M. Ammon, M. Baggioli, A. Jim\'enez, O. Pujol\`as, ``Black hole elasticity and gapped transverse phonons in holography", JHEP 1801(2018)129, 	arXiv:1708.08477 [hep-th].
 \bibitem{Alberte:1711}
L. Alberte, M. Ammon, M. Baggioli, A. Jim\'enez, O. Pujol\`as, ``Holographic Phonons", Phys. Rev. Lett. 120, 171602 (2018), arXiv:1711.03100 [hep-th].
 \bibitem{Amoretti:1711}
A. Amoretti, D. Are\'an, B. Gout\'eraux, D. Musso, ``Effective holographic theory of charge density waves", Phys. Rev. D 97, 086017 (2018), arXiv:1711.06610 [hep-th].
 \bibitem{Amoretti:1712}
A. Amoretti, D. Are\'an, B. Gout\'eraux, D. Musso, ``DC resistivity of quantum critical, charge density wave states from gauge-gravity duality", Phys. Rev. Lett. 120, 171603 (2018), arXiv:1712.07994 [hep-th].
\bibitem{Poovuttikul:1801}
S. Grozdanov, N. Poovuttikul, ``Generalised global symmetries in states with dynamical defects: the case of the transverse sound in field theory and holography", Phys. Rev. D 97, 106005 (2018), arXiv:1712.07994 [hep-th].
 \bibitem{Baggioli:1805}
 M. Baggioli, A. Buchel, ``Holographic Viscoelastic Hydrodynamics", arXiv:1805.06756 [hep-th].
  \bibitem{Kuang:1808}
G. Filios,  P. A. Gonz\'alez, X.-M. Kuang, E. Papantonopoulos, Y. V\'asquez, ``Spontaneous Momentum Dissipation and Coexistence of Phases in Holographic Horndeski Theory", arXiv:1808.07766 [hep-th].
 \bibitem{Amoretti:1812}
 A. Amoretti, D. Are\'an, B. Gout\'eraux, D. Musso, ``A holographic strange metal with slowly fluctuating translational order", arXiv: 1812.08118 [hep-th] [hep-th].
 \bibitem{Rozali:1211}
 M. Rozali, D. Smyth, E. Sorkin, J. Stang, ``Holographic Stripes", Phys. Rev. Lett. 110, 201603 (2013), arXiv:1211.5600 [hep-th].
 \bibitem{Donos:1303}
 A. Donos, JHEP 1305, 059 (2013) [arXiv:1303.7211 [hep-th]; A. Donos, J. Gauntlett, Phys. Rev. D 87, 126008 (2013), arXiv:1303.4398 [hep-th].
  \bibitem{Ling:1404}
 Y. Ling, C. Niu, J.-P. Wu, Z. Xian, H. Zhang, ``Metal-insulator Transition by Holographic Charge Density Waves", Phys. Rev. Lett. 113, 091602 (2014), arXiv:1404.0777 [hep-th].
   \bibitem{lili:1612}
S. Cremonini, L. Li, J. Ren, Phys.Rev. D95 (2017) no.4, 041901,  arXiv:1612.04385 [hep-th]; S. Cremonini, L. Li, J. Ren,  JHEP 1708 (2017) 081, arXiv:1705.05390 [hep-th].
\bibitem{lili:1706}
R.-G. Cai, L. Li, Y.-Q. Wang, J. Zaanen, ``Intertwined Order and Holography: The Case of Parity Breaking Pair Density Waves", Phys. Rev. Lett. 119 (2017) no.18, 181601, arXiv:1706.01470 [hep-th].
 \bibitem{Jokela:1708}
N. Jokela, M. Jarvinen, M. Lippert, ``Holographic pinning", Phys. Rev. D 96, 106017 (2017),  arXiv:1708.07837[hep-th].
 \bibitem{Donos:1801}
A. Donos, J. Gauntlett, T. Griffin, V. Ziogas, ``Incoherent transport for phases that spontaneously break translations", JHEP 1804(2018)053, arXiv:1801.09084 [hep-th].
 \bibitem{Gouteraux:1803}
B. Gout\'eraux, N. Jokela, A. P\"onni
, ``Incoherent conductivity of holographic charge density waves", JHEP 1807 (2018) 004, arXiv:1803.03089  [hep-th].
 \bibitem{musso}
Daniele Musso, ``Simplest phonons and pseudo-phonons in field theory", arXiv:1810.01799 [hep-th].
 \bibitem{vegh}
D. Vegh, Holography without translational symmetry, arXiv:1301.0537  [hep-th].
 \bibitem{WJL:1602}
B. Gout\'eraux, E. Kiritsis, W.-J. Li,
``Effective holographic theories of momentum relaxation and violation of conductivity bound", JHEP 1604 (2016) 122, arXiv:1602.01067[hep-th].
 \bibitem{Andrade:1311}
T. Andrade and B. Withers, ``A simple holographic model of momentum relaxation", JHEP 05 (2014) 101, arXiv:1311.5157 [hep-th].
 \bibitem{Gouteraux:1401}
B. Gout\'eraux, ``Charge transport in holography with momentum dissipation", JHEP 04 (2014) 181, arXiv:1401.5436 [hep-th].
 \bibitem{Kim:1501}
K.-Y. Kim, K. K. Kim, M. Park, ``A Simple Holographic Superconductor with Momentum Relaxation", JHEP 04 (2015) 152, arXiv:1501.00446 [hep-th].
 \bibitem{Gruner}
G. Gr\"uner, ``The dynamics of charge density waves", Rev. Mod. Phys. 60. 1129.
 \bibitem{baggioli:1411}
M. Baggioli, O. Pujol\`as, ``Electron-Phonon Interactions, Metal-Insulator Transitions, and Holographic Massive Gravitys", Phys. Rev. Lett. 114, 251602 (2015),  arXiv:1411.1003 [hep-th].
 \bibitem{Kadanoff:1963}
L. P. Kadanoff and P. C. Martin, ``Hydrodynamic equations and correlation
functions", Annals of Physics 24 (1963) 419-469.
 \bibitem{tomas:1710}
T. Andrade, A. Krikun, K. Schalm, J. Zaanen, and P. C. Martin, ``Doping the holographic Mott insulator", Nature Physics 14, 1049-1055 (2018), arXiv:1710.05791 [hep-th].
  \bibitem{tomas:1512}
T. Andrade, A. Krikun, ``Commensurability effects in holographic homogeneous lattices", JHEP 05 (2016) 039, arXiv:1512.02465 [hep-th].
 \bibitem{Baggioli:1807}
 M. Baggioli, K. Trachenko, ``Solidity of liquids: How Holography knows it", arXiv:1807.10530 [hep-th].
  \bibitem{Baggioli:1808}
 M. Baggioli, K. Trachenko, ``Maxwell interpolation and close similarities between liquids and holographic models ", arXiv:1808.05391 [hep-th].
 \bibitem{Baggioli:up}
 M. Baggioli, V. Castillo, O. Pujol\`as,  S. Petel, to appear.
  \bibitem{WJL:up}
 M. Baggioli, W.-J. Li, J.-P. Wu, in preparation.
\bibitem{Lindgren:2015lia}
  J.~Lindgren, I.~Papadimitriou, A.~Taliotis and J.~Vanhoof,
  ``Holographic Hall conductivities from dyonic backgrounds,''
  JHEP {\bf 1507}, 094 (2015)
  %doi:10.1007/JHEP07(2015)094
  [arXiv:1505.04131 [hep-th]].
%\cite{Caldarelli:2016nni}
\bibitem{Caldarelli:2016nni}
  M.~M.~Caldarelli, A.~Christodoulou, I.~Papadimitriou and K.~Skenderis,
  ``Phases of planar AdS black holes with axionic charge,''
  JHEP {\bf 1704}, 001 (2017)
 % doi:10.1007/JHEP04(2017)001p
  [arXiv:1612.07214 [hep-th]].
  \bibitem{Skenderis}
  S. Skenderis, ``Lecture Notes on Holographic Renormalization ", Class. Quant. Grav. 19: 5849-5876, 2002, arXiv:0209067 [hep-th].
 %\cite{Lindgren:2015lia}
\end{thebibliography}
\end{document}